\documentstyle[12pt]{article}
\title{\bf On Signature Transition and Compactification in Kaluza-Klein Cosmology }
\author{F. Darabi\thanks{e-mail: f-darabi@cc.sbu.ac.ir} $\:\:$
and
$\:$ H. R. Sepangi\thanks{e-mail: hr-sepangi@cc.sbu.ac.ir}
\\
{\small Department of Physics, Shahid Beheshti University, Evin, Tehran 19839, Iran.}
\\
}
\begin{document}
\maketitle
\vspace{15mm}
\begin{abstract}
We consider an empty (4+1) dimensional Kaluza-Klein universe with a negative cosmological
constant and a Robertson-Walker type metric. It is shown that the solutions
to Einstein field equations have
degenerate metric and exhibit transitioins from a Euclidean to a Lorentzian
domain. We then suggest a mechanism,
based on signature transition which leads to compactification
of the internal space in the Lorentzian region as $a \sim |\Lambda|^{1/2}$.
With the assumption of a very small value
for the cosmological constant  we find  that  the size of the
universe $R$ and the internal scale factor $a$ would be related according to
$Ra\sim 1$ in the Lorentzian region.
The corresponding Wheeler-DeWitt equation has exact solution in
the mini-superspace giving rise to a quantum state which peaks in the
vicinity of the classical solutions undergoing signature transition.
\end{abstract}

\vspace{15mm} 

\newpage

\section{Introduction}

The question of signature transition in classical and quantum cosmological
models has been of some interest in the past few years. It was first
addressed in the work of Hartle and Hawking \cite{H} in which they examined
quantum cosmologies admitting quantum amplitudes in the form of a sum over
all compact Riemannian manifolds whose boundaries coincide with the loci of
signature change. Other workers have also investigated this question and
considered signature transition in general relativity by adopting model
theories which mostly rely on Einstein's field equations coupling to a
scalar field \cite{all}. The solution of the resulting field equations under
a properly parametrized metric, when interpreted suitably, would then indicate
a change of signature.

A particular model, relevant to the present work, is that of Dereli and
Tucker \cite{DT} in which a self interacting scalar field is coupled to
Einstein's field equations with a potential containing a Sinh-Gordon scalar
interaction. These equations are then solved exactly for the scalar field
and the
scale factor as dynamical variables, giving rise to cosmological solutions
with a degenerate metric, describing a continuous signature transition from
a Euclidean domain to a Lorentzian space-time in a spatially flat
Robertson-Walker cosmology. The corresponding quantum cosmology has also
been investigated \cite{DOT} where the Wheeler-DeWitt equation arises from
an anisotropic oscillator-ghost-oscillator vanishing Hamiltonian. It is
solved exactly and leads to normalizable states with the quantum states
constructed as belonging to distinct Hilbert subspaces each of which being
characterized by a particular ``quantization'' condition on the parameters
of the scalar field potential. These quantum states correspond to excited
quantum cosmologies and relate to classical solutions without resort to WKB
approximation techniques. A similar analysis of the classical and quantum
theory of a scalar (dilaton) field interacting with gravity has been
reported in two dimensions \cite{OT} in which a class of analytic solutions
to the Wheeler-DeWitt equation relate, in a remarkable way, to the general
solution of the classical field equations.

In this paper, we consider a (4+1) dimensional Kaluza-Klein cosmology with a
{\it negative} cosmological constant and a
Robertson-Walker type metric having two dynamical variables, the usual scale
factor $R$ and the internal scale factor $a$. Following \cite{DT} and
\cite{DOT}, we insist on a preferred coordinate that controls the evolution
of signature dynamics, seeking suitably smooth continuous solutions for
$R$ and $a$
passing through the hypersurface of signature change.
These classical solutions admitting signature transition  would then
suggest a compactification mechanism for the internal scale
factor $a$.
Some authors \cite{GK, EMB, GSV} have already considered various compactification
mechanisms in different models. In particular in \cite{EMB}, the author has
suggested a compactification mechanism based on signature change for a {\it positive}
cosmological constant. Here, we will discuss
the differences and similarities between these
models and the one presented here
in order to see how the results can be compared.
We then find exact solutions to the
corresponding Wheeler-DeWitt equation arising from an isotropic
oscillator-ghost-oscillator vanishing Hamiltonian. Due to this isotropy,
there is no
``quantization'' condition and thus no excited cosmologies. However, we show
that the desired quantum state, identified with a non-dispersive
wave packet, relates to classical solutions in that it peaks
in the vicinity of the classical
loci corresponding to this cosmology.

\section{Classical Cosmology}

We start with the metric considered in \cite{D} in which the space-time is
assumed to be of Robertson-Walker type having a compactified space which is
assumed to be the circle $S^1$. In this paper we adopt the real chart $\{\beta, r^1, r^2,
r^3, \rho\}$ with $\beta$, $r^i$ and $\rho$ denoting the lapse function, the
space coordinates and the compactified space coordinate respectively. We
therefore take
\begin{equation}
d s^2 = -\beta {d\beta}^2+{\bar{R}}^2 (\beta)\frac{dr^i \: dr^i}{(1+
\frac{k r^2}{4})^2}+ {\bar{a}}^2(\beta) {d\rho}^2 ,\:\:\:i=1, 2, 3
\label{1}
\end{equation}
where $k=0,\pm 1$ and $\bar{R}(\beta)$ is the scale factor and $\bar{a}(\beta)$ is the
radius of the compactified space, both of which are assumed to depend only
on the lapse function $\beta$. The signature of the metric is Lorentzian for
$\beta>0$ and Euclidean for $\beta<0$. For positive values of $\beta$ (Lorentzian region),
one can recover the cosmic time by writing $t=\frac{2}{3}\beta^{\frac{3}{2}}$,
leading to
\begin{equation}
d s^2 = -{dt}^2+R^2 (t) \frac{dr^i \: dr^i}{(1+\frac{k r^2}{4})^2}+a^2 (t)
{d\rho}^2
\end{equation}
where $R(t)=\bar{R}(\beta(t))$ and $a(t)=\bar{a}(\beta(t))$ in the $\{t,
r^i, \rho\}$ chart. As in \cite{DT}, we formulate our differential
equations in a region that does not include $\beta=0$ and seek
real solutions for $R$ and $a$ smoothly passing through the
$\beta=0$ hypersurface. The curvature scalar corresponding to metric (\ref{1})
is obtained as
\begin{equation}
{\cal R}=6 \left[\frac{\ddot{R}}{R}+\frac{k+{\dot{R}}^2}{R^2}\right] +2\frac{
\ddot{a}}{a}+ 6 \frac{\dot{R}}{R} \frac{\dot{a}}{a}
\end{equation}
where a dot represents differentiation with respect to $t$. Substituting
this result into Einstein-Hilbert action with a cosmological constant
$ \Lambda$
\begin{equation}
I=\int\!\sqrt{-g} ({\cal R}-\Lambda) dt\:d^3 r\:d\rho
\label{4}
\end{equation}
and integrating over spatial dimensions gives an effective Lagrangian $L$ in the
mini-superspace ($R$,$a$) as

\begin{equation}
L=\frac{1}{2} R a {\dot{R}}^2+\frac{1}{2}R^2 \dot{R}\dot{a}-\frac{1
}{2}k R a+\frac{1}{6} \Lambda R^3 a.
\label{5}
\end{equation}

\section{Solutions}

By defining $\omega ^2\equiv -\frac{2\Lambda }3$ and changing the variables
as \cite{W}
\begin{equation}
u=\frac 1{\sqrt{8}}\left[ R^2+Ra-\frac{3k}\Lambda \right] ,
\hspace{10mm}v=\frac 1{\sqrt{8}}\left[ R^2-Ra-\frac{3k}\Lambda \right]
\label{6}
\end{equation}
$L$ takes on the form
\begin{equation}
L=\frac{1}{2} \left[({\dot u}^2-\omega ^2u^2)-({\dot v}^2-\omega ^2v^2)\right].
\label{7}
\end{equation}
A point $(u,v)$ in this mini-superspace represents a 4-geometry.
The classical equations of motion are given by
\begin{equation}
\ddot u=-\omega ^2u,\hspace{10mm}\ddot v=-\omega ^2v.
\label{8}
\end{equation}
Choosing the initial conditions $\dot u(0)=\dot v(0)=0$
, the solutions are obtained as
\begin{equation}
u(t)=A\cosh \left( \sqrt{\frac{2\Lambda }3}t\right) ,\hspace{10mm}
v(t)=B\cosh \left( \sqrt{\frac{2\Lambda }3}t\right)
\label{9}
\end{equation}
where $A$ and $B$ are constants to be determined later. Assuming the full
(4+1) dimensional Einstein equations to hold, this implies that the
Hamiltonian corresponding to $L$ in (\ref{7}) must vanish, that is
\begin{equation}
H=\frac{1}{2} \left[({\dot u}^2+\omega ^2u^2)-({\dot v}^2+\omega^2v^2)\right]=0
\label{10}
\end{equation}
which describes an isotropic oscillator-ghost-oscillator system. Now, the
solutions (\ref{9}) must satisfy the constraint of vanishing Hamiltonian.
Thus, substitution of equations (\ref{9}) into (\ref{10}) gives a relation
between the constants $A$ and $B$
\begin{equation}
\label{11}A=\pm B
\end{equation}
implying that we can rewrite the solutions (\ref{9}) as
\begin{equation}
u(t)=A\cosh \left( \sqrt{\frac{2\Lambda }3}t\right) ,\hspace{10mm}
v(t)=\epsilon A\cosh \left( \sqrt{\frac{2\Lambda }3}t\right)
\label{12}
\end{equation}
where $\epsilon $ takes the values $\pm 1$ according to the choices in (\ref{11}).
Classical solutions (\ref{12}) may be displayed as trajectories $u=\pm v$ in the
mini-superspace ($u,v$).
We recover $R(t)$ and $a(t)$ from $u(t)$ and $v(t)$ as
\begin{equation}
\label{13}
\begin{array}{ll}
R(t)=\left[ \sqrt{2}( u(t)+v(t) )+\frac{3k}{\Lambda} \right] ^{1/2}  \\
 \\
a(t)=\left[ \sqrt{2}( u(t)+v(t) )+\frac{3k}{\Lambda} \right] ^{-{1/2}}\left[
\sqrt{2} ( u(t)-v(t) )\right] .
\end{array}
\end{equation}
For $\epsilon =-1$ one finds the solutions in terms of $\beta$ as
\begin{equation}
\begin{array}{ll}
R_c=\sqrt{\frac{3k}{\Lambda} } &  \\
&  \\
a(\beta)=\sqrt{\frac{\Lambda}{3k}}\cosh \left( \sqrt{\frac{2\Lambda}{3}}
\frac{2}{3}\beta^{3/2} \right)
\label{14} &
\end{array}
\end{equation}
where $A=\frac 1{\sqrt{8}}$ is taken for convenience\footnote{
Note that the dimension of $A$ is (length)$^2$.} and solutions (\ref{12})
have been used. Also for $\epsilon =+1$, one finds the solutions
\begin{equation}
\begin{array}{ll}
R(\beta)=\left[\cosh \left(\sqrt{\frac{2\Lambda}{3}} \frac{2}{3}
\beta^{3/2}\right)+\frac{3k}{\Lambda}\right]^{1/2} &  \\
&  \\
a=0.
\label{15} &
\end{array}
\end{equation}

\section{Signature Transition and Compactification}

In this section we discuss signature transition occurring in the model
presented above and show that it induces compactification on the internal space.
However, before doing so we give a brief history of the mechanisms of
compactification related to the present work, in particular the recent works
discussed in \cite{GK} and \cite{EMB}.

In \cite{EMB}, it is shown that for a higher dimensional FRW model
with $S^3 \times S^6$
as spatial sections with two scale factors $a_1, a_2$ and a
positive cosmological constant, the classical
signature change induces compactification. This is done by a
dynamical mechanism that drags
the size of $S^6$ down and gives rise to a long-time stability at
an unobservably small
scale. This mechanism is based on the existence of a signature transition
and the
interplay between the causal structure of the Wheeler-DeWitt
metric and the sign of the corresponding potential $W$ appearing in
the action defined in the mini-superspace ($a_1,a_2$). In order to
diagonalize the kinetic term in the action
a new set of variables ($u,v$) are defined. In this new mini-superspace
the curve satisfying $W=0$ consists of two branches which are
connected smoothly. In one branch,
the compactification occurs for $S^6$ and in
the other, it occurs for $S^3$. For the compactification of $S^6$, one
finds $a_1 (t)$ oscillating around some
linearly growing average, whereas $a_2 (t)$
performs damped oscillations around a limiting value of
order $\Lambda^{-\frac{1}{2}}$,
both solutions being stable against small perturbations.
The effective five-dimensional space-time metric, obtained by taking the
proper time average, has a
Lorentzian signature, undergoes exponential inflation
(in $\Lambda$) in $S^3$ and induces compactification on $S^6$ of order
$\Lambda^{-\frac{1}{2}}$. This requires a large cosmological constant
in order to be consistent with the unobservability
of the compactified dimensions.
On the other hand, stopping inflation requires switching off
$\Lambda$, but then the radius of the compactified $S^6$ will
blow up. Usually, this type of problem is expected in the
presence of a large positive cosmological constant whose
possible solutions are suggested in \cite{EMB}.

In \cite{GK} however, the compactification mechanism is
studied for a $D+1$ dimensional
toroidally compact Kaluza-Klein cosmology with a negative
cosmological constant consisting of matter that is either dust
or coherent excitations of a dilaton field. In this model, compactification
is done by diagonalizing the classical Hamiltonian which leads
to a new $D$-dimensional mini-superspace. Applying the
Heisenberg equations of motion and taking the expectation values,
one finds the cosmic time dependence of the expectation values
of the mini-superspace variables. It is then shown that the expectation
values of some of the dimensions show a quantum inflationary phase
while simultaneously the remaining dimensions show a quantum
deflationary phase giving rise to compactification. In this model, the eternal
inflation-compactification is a problem whose solution is
based on the tunneling of the negative cosmological constant to
zero in the context of a dilaton field model
having a potential with two local minima. The quantum
inflation-deflation (QID) era is then realized by oscillations of
the dilaton field around the absolute minimum
(where $\Lambda<0$). After tunneling to $\Lambda\simeq 0$
this quatum phase disappears allowing a classical description at later times.

There are differences and similarities between the models described above
and the one presented in this paper.
In our model, the metric is a 5-dimensional Kaluza-Klein with
a negative cosmological constant whereas in
\cite{EMB}, the metric is 10-dimensional with a cosmological constant
which is positive. However, both models use signature transition as
the process for addressing compactification, but
through two completely different mechanisms. As for the model presented in
\cite{GK}, we find a similarity in their
assumption of a negative cosmological constant, but their compactification
mechanism and matter contents are very different. They use either
dust or  a dilaton field as the matter content, contrary to our model in which
there is no matter.
However, in spite of the above differences, the behaviour of the two scale
factors $R$ and $a$ in the present model merits some discussion under various
possible choices of $\Lambda$ in order to compare them with
that of \cite{GK, EMB} at the formal level. To this end, we first discuss
signature transition in the model presented here by  seeking suitably
smooth continuous solutions for $R$ and $a$
passing through the hypersurface of signature change $\beta=0$.

The classical solutions (\ref{14}, \ref{15})
describe an empty Kaluza-Klein universe with a {\it negative}
cosmological constant.
When $ \epsilon =-1$, the universe takes the same
constant scale factor $R_c$ in both Euclidean and Lorentzian regions, hence
it is continuous at $\beta=0$. The $\beta $ dependent scale factor
$a(\beta )$ is unbounded in the Euclidean region
$\beta<0$, passing continuously through $\beta=0$ and exhibiting
bounded oscillations in the
Lorentzian region $\beta >0$. The reality conditions on $R_c$
and $a(\beta )$ forces $k$ to be
negative, thus rendering this universe open with $k=-1$
\footnote{The case $k=0$ would give rise to a divergent $a$ and zero $R$.}.
For $|\Lambda|\gg0$ the solution (\ref{14}) will give rise to a small
constant scale factor $R$ passing continuously from Euclidean to
Lorentzian regions. The scale factor $a$ will be enormously
large in both regions compared to the scale factor $R$ and passes
through $\beta=0$ continuously. The scale factor
$R$ is now compactified to a small size of order $|\Lambda|^{-\frac{1}{2}}$.
Taking $|\Lambda| \simeq 0$, it is seen that for $\beta
\ll 0$, $a(\beta )$ can become large and for $\beta \leq 0$, it tends to
become very small in comparison with $R_c$.
It passes through $\beta=0$
continuously with a very small value $a(0)=R_c^{-1}$ and oscillates for $\beta >0$
 with amplitude $a(0)$ and a varying frequency. One of the most
interesting features in this case is that signature transition now induces
compactification on the scale factor $a$ in the Lorentzian region, dragging it to a small
size of order $|\Lambda|^{\frac{1}{2}}$. The first
zero of this oscillatory function in the Lorentzian region occurs at
$$
\beta _0=\left(\frac{3\pi }4\sqrt{\frac 3{2|\Lambda |}}\right)^{2/3}.
$$
It is seen that the smallness of the
cosmological constant (large $\beta_0$) allows for an extended
Lorentzian region  $0\leq\beta< \beta_0$ which would
correspond to a Kaluza-Klein cosmology with a large scale factor $R$ and
a stable compactified scale factor $a$, see figure 1. The long-time
stability of the compactification is verified for the present bound on the
cosmological constant $|\Lambda| \sim 10^{-56}$ cm$^{-2}$
since  then $\beta_0\geq$  {\it present age of universe}.

When $\epsilon =+1$, the solutions (\ref{15}) describe
an empty Kaluza-Klein universe for which the internal space has zero
size. Assuming a negative cosmological constant $\Lambda= 3k$ with $k=-1$,
the solution $R(\beta)$ is real and behaves exponentially for $\beta<0$,
passing through $\beta=0$ continuously and oscillating for $\beta>0$.
This, indeed, compactifies $R$ to a small size $0\leq R \leq \sqrt{2}$
in the Lorentzian region. Taking $|\Lambda|\simeq 0$ leads to
a real scale factor $R(\beta)$. It has large values at both regions, behaving
exponentially for $\beta \ll 0$, tending to a large constant value
$\sim |\Lambda|^{-1/2}$ for $\beta \leq0$, passing through $\beta=0$
continuously and oscillating about this large value for $\beta>0$.
There is good agreement between $R$ and its present observational bound
$R \sim 10^{28}$ cm for the choice $|\Lambda| \sim 10^{-56}$ cm$^{-2}$.

In conclusion, from the discussion given above, we find that
choosing a small cosmological constant would lead to a good agreement
between the size of the universe  arising from
the solutions (\ref{14}, {15}) and its
present observational bound. This choice of $\Lambda$ also leads
to an  agreement between
the present unobservability of the size of the compactified dimension $a$
and that resulting  from the above solutions.
As our solutions (\ref{14}) show, we may recognize two hierarchial phases in
the Lorentzian region as
\begin{eqnarray}
t\simeq0 \:,\: |\Lambda|\gg0 \:\: \Rightarrow R \sim |\Lambda|^{-1/2}
\:,\: a \sim |\Lambda|^{1/2}  \nonumber
\end{eqnarray}
and
\begin{eqnarray}
0<t<t_0 \:,\: |\Lambda|\simeq0 \:\: \Rightarrow R \sim |\Lambda|^{-1/2}
\:,\: a \sim |\Lambda|^{1/2}  \nonumber
\end{eqnarray}
both exhibiting the same relation $Ra \sim1$ ($t_0$ being the present age of universe).
Thus, we have found
a classical link between the size
of the extra dimension $a$ and that of the visible dimension $R$ in the Lorentzian region.
This relation may account for the existence
of a transformation
$\Lambda \rightarrow \Lambda^{-1}$ leading to duality transformations
$R \rightarrow a \sim R^{-1}$ and $a \rightarrow R \sim a^{-1}$.
Therefore, such dualities
may be studied in the context of a theory in which
a large $|\Lambda|$ in the
very early universe will result in a  small $|\Lambda|$
at later times and hence an initial small  $R$
becomes very large while simultaneously the initial large
$a$ compactifies to a very small size.
Therefore, if we regard the relation $Ra\sim 1$ as the main characteristic
of this duality theory then switching off the large
cosmological constant would be a natural result in order to be consistent
with present observations.

A clear similarity is seen between our model and that of \cite{GK}
in the sector of internal space compactification.
Both of these models predict
a deflationary phase for compactification of the internal dimensions
before the onset of classical evolution. Considering  equation
(\ref{14}) with $|\Lambda|\simeq0$ and as is seen in figure 1, there
is an exponentially decreasing $a(\beta)$ in the Euclidean
region $-\infty<\beta\leq0$ after which it undergoes a long-time
stability in the Lorentzian
region $0\leq\beta<\beta_0$. Usually, the  classical evolution
of the universe begins at $\beta=0$ and is considered to be in the
Lorentzian sector so that the Euclidean sector may be assumed to be a
pre-classical era \cite{all}. Therefore, in the context
of the present signature changing approach to compactification of
the internal space $a$ we also find a pre-classical era (Euclidean region)
where a deflationary phase occurs after which the classical
evolution begins. One may also resort to a mechanism in a more fundamental
theory which would lead to a rapid deflationary phase (a large $\Lambda$) and
then a long time stable phase (a small $\Lambda$) for $a(\beta)$.
In fact, some switching off mechanisms could be proposed by which $\Lambda$
could relax to zero.

An alternative similarity to \cite{GK} will be obtained
if we consider our model as an effective part of a more fundamental
theory in which a negative cosmological
constant $\Lambda \sim 3k$ can tunnel to a very small value.
Then, considering the Lorentzian solutions (\ref{14}) with $\beta>0$,
and assuming $\Lambda \sim 3k$ we have
two equally compact sizes $a \sim R$ initially at $t\simeq0$.
After the tunneling of
$|\Lambda|$, the scale factor $R$ becomes very large and  $a$ gets very
small, both evolving classically thereafter.
This picture may be identified, at least
at the formal level, with
the (QID) model in which, at first, all dimensions have equally
compact sizes but
after the (QID) phase ends, some of them become larger and the
remaining ones get smaller. The cosmological constant can then
tunnel from a large negative
value to zero and the universe can be described classically afterwards.

Comparison with \cite{EMB} shows that in our mini-superspace
$(u,v)$, one  finds, as in \cite{EMB}, two branches $u=\pm v$
corresponding to the vanishing
of potential $W=\frac{1}{2} \omega^2(u^2 - v^2)=0$, c.f. equation (\ref{7}).
The branch $u=-v$ of the curve $W=0$ gives the compactification
either of $R$ or $a$ according to the choice of the heirarchial
cosmological constants. The other branch $u=+v$ leads to the compactification
of $a$ to zero  and $R$ to $0 \leq R \leq \sqrt{2}$.

As before, of particular interest is the branch $u=-v$ with a very small
negative (magnitude) cosmological constant in which the size of
compactification of the internal scale factor
$a$, in the Lorentzian region, is of order $|\Lambda|^{\frac{1}{2}}$.
At first glance, this result seems to be in conflict with that of \cite{EMB}
where $a\sim \Lambda^{-\frac{1}{2}}$ (for positive $\Lambda$), but since
the cosmological constant in \cite{EMB} is assumed to be large and
the one here is small, we see  that the results are
in agreement
\footnote{To see this, one may compare the long-time behaviour
and sizes of $a$ in the
Lorentzian sector of figure 1 here, and figure 2 in \cite{EMB}.}.
The only difference between the result of the compactification mechanisms
in \cite{EMB} and that proposed here concerns the scale factor $R$. In \cite{EMB},
the size of $S^3$ undergoes an eternal inflation
whereas the scale factor $R$ is constant here and in  agreement with its observational bound.

\section{Quantum Cosmology}

One of the most interesting topics in the context of quantum cosmology is the
mechanisms through which  classical cosmology may emerge from quantum theory.
When does a Wheeler-DeWitt
wave function predict a classical space-time? Indeed, any attempt in
constructing a viable quantum gravity requires understanding
the connections between classical and quantum
physics. Much work has been done in this direction over the past decade.
Actually, there is some tendency towards using semiclassical
approximations in dividing the behaviour of the wave
function into two types, oscillatory or exponential which are
supposed to correspond to classically allowed or forbiden regions.
Hartle \cite{Hartle} has put forward a simple rule for applying quantum mechanics to
a single system (universe): If the wave function is
sufficiently peaked about some region in the configuration space we
predict to observe a correlation between the observables which
characterize this region. Halliwell \cite{Hall} has shown that the oscillatory
semiclassical WKB wave function is peaked about a region of the
mini-superspace in which the correlation between the coordinate
and momentum holds good and stresses that both $correlation$ and
$decoherence$ are necessary before one can say a system is classical.
Using Wigner functions, Habib and Laflamme \cite{Hab} have studied the mutual
compatibility of these requirements and shown that
some form of coarse graining is necessary  for classical prediction
from WKB wave functions. Alternatively, Gaussian or coherent
states with sharply peaked wave functions
are often used to obtain classical limits by constructing
wave packets.

A new aspect arises in quantum cosmology if the wave packets are to
be constructed which would trace out a classical trajectory.
In this case, the normalizability of the
wave function is needed in order to have a correlation between classical
and quantum cosmology. Of course, because of the superposition principle,
some interference between the coherent states exist but enlarging
the configuration space by adding a large number of higher
degrees of freedom interacting with the mini-superspace
variables leads to a decoherence effect.

Recently, this aspect of correspondence between classical and
quantum cosmology has become of interest \cite{kiefer}, especially
in the context of signature transition \cite{DOT,OT}.
In \cite{DOT}, the authors have exactly solved the Wheeler-DeWitt equation in the
form of an anisotropic oscillator-ghost-oscillator constraint
and constructed states that highlight the classical trajectories and
admit signature transition without
resort to WKB approximations, hence avoiding the decoherence problem.
Also in \cite{Lidsey}, it is shown that
a large subset of generalized two-dimensional dilaton gravity models
are dynamically equivalent to the isotropic oscillator-ghost-oscillator
constraint and there may be correspondence
between classical and quantum configurations like that obtained in \cite{OT}.

In this section we shall concentrate on those aspect of
correspondence between classical and quantum cosmology which are studied
in \cite{DOT}.
The Lagrangian in (\ref{7}) describes a classical Kaluza-Klein cosmology.
The corresponding quantum cosmology is described by the Wheeler-DeWitt
equation resulting from Hamiltonian (\ref{10}) and can be written as
\begin{equation}
\left[ \frac{\partial ^2}{\partial u^2}-\frac{\partial ^2}{\partial v^2}
-(u^2-v^2)\omega ^2\right] \Psi (u,v)=0
\end{equation}
where $\omega ^2\equiv -\frac{2\Lambda }3$.
This is an isotropic oscillator-ghost-oscillator constraint.
For $\Lambda <0$, the ``zero
energy'' solutions belong to a subspace of the Hilbert space spanned by
separable eigenfunctions of a 2-dimensional isotropic simple harmonic
oscillator Hamiltonian, and can be written as
\begin{equation}
\Phi _{(n_1,n_2)}(u,v)=\alpha _{n_1}(u)\beta _{n_2}(v)\hspace{10mm}
n_1,n_2=0,1,2,\ldots
\end{equation}
with
\begin{equation}
\alpha _n(u)\equiv (\frac \omega \pi )^{1/4}\frac{e^{-\frac{\omega u^2}2}}
{\sqrt{2^nn!}}H_n(\sqrt{\omega }u),
\end{equation}
\begin{equation}
\beta _n(v)\equiv (\frac \omega \pi )^{1/4}\frac{e^{-\frac{\omega v^2}2}}
{\sqrt{2^nn!}}H_n(\sqrt{\omega }v)
\end{equation}
provided $(n_1+\frac 12)\omega =(n_2+\frac 12)\omega $ with the restriction
$n_1=n_2 \equiv n$
\footnote{Note that in \cite{DOT} there is a quantization condition on $\omega$
resulting in distinct Hilbert subspaces corresponding to excited cosmologies.
}. $H_n(x)$ is the Hermit polynomial and the eigenfunctions are normalized
according to
\begin{equation}
(\alpha _n,\alpha _m)=\delta _{n,m},\hspace{5mm}(\beta _n,\beta _m)=\delta
_{n,m}
\end{equation}
where ( , ) denotes the inner product. The solutions $\Phi _{(n,n)}(u,v)$
span a Hilbert subspace of measurable square integrable functions on $R^2$
in the form
\begin{equation}
\label{23}\Psi (u,v)=\sum_{n=0}^\infty c_n\Phi _{(n,n)}(u,v)
\end{equation}
where $c_n\in C$ and
\begin{equation}
(\Phi _{(n,n)},\Phi _{(n^{\prime },n^{\prime })})=\delta _{n,n^{\prime }}.
\end{equation}
We are interested in constructing a coherent wave packet with
good asymptotic behaviour in the mini-superspace, peaking in
the vicinity of one of the classical loci $u=\pm v$
in the $\{u,v\}$ configuration space. It is well-known that for a harmonic
oscillator, non-dispersive wave packets may
be constructed by superposition of energy eigenfunctions.
Therefore, the wave packet (\ref{23}) is what we need.
We take the solution as being represented by equation (\ref{23}) with the
sum to be truncated at a suitable value of $n$ displaying this peak. Figures
2 and 3 show surface and density plots of $|\Psi (u,v)|^2$, where we have
taken the combinations from 6 terms with all $c_n$ up to $c_5$ taken to be
unity. Taking more terms would only have a small effect on the results. It
is seen that a good correlation exists between these patterns and the
classical loci $u= v$ in the configuration space $\{u,v\}$.

Wave packets
in the mini-superspace can only be understood as unparametrized tubes.
This is because the Wheeler-DeWitt equation does not have an intrinsic
time parameter. However, in order to relate the properties of the wave
packet (\ref{23}) to an evolving classical
cosmology  with classical time in the Lorentzian region, one
may take $\{\beta\}$ as a family of coordinate functions labeling
some subset of the real line containing the point $\beta=0$.
Then the loci $u=\pm v$ admit
parametrizations in terms of
$\{\beta\} $
\begin{equation}
u(\beta)=A\cosh \left( \sqrt{\frac{2\Lambda }{3}}\frac{2}{3}\beta^{\frac{3}{2}
}\right) ,\hspace{10mm}
v(\beta)=\epsilon A\cosh \left( \sqrt{\frac{2\Lambda }{3}}\frac{2}{3}
\beta^{\frac{3}{2}}\right)
\end{equation}
with the point $\beta =0$ implying a transition
from Euclidean to Lorentzian signature.
Now, a change of coordinate $\beta \rightarrow \beta'= F(\beta)$ will induce
a change of parametrization for the classical loci and, for $\beta>0$
 with $F$ monotonic,
would correspond to an alternative choice of classical time. To the extent
that the
classical loci are highlighted by the state (\ref{23}) we say that the family
of classical times, regarded as alternative parametrization of such loci,
arise dynamically from this state.

\section{Conclusions}

In this paper we have considered the
solutions of Einstein equations in an empty (4+1) dimensional
Kaluza-Klein cosmology with a Robertson-Walker type metric having
a negative cosmological constant. These solutions admit a degenerate
metric signifying a signature transition
from Euclidean to Lorentzian domains. Motivated by the subject of compactification
which has been studied in the context of signature change for a positive
cosmological constant, we have shown that here, signature
transition can provide a compactification either for the usual scale
factor $R$ or the internal scale factor $a$ according to the
choice of hierarchial negative cosmological constant. The most interesting
and desirable case emerges by taking a  very small
negative cosmological constant which would then predict
the relation $Ra \sim 1$ and compactification of
$a$ as $a \sim |\Lambda|^{1/2}$ for the Lorentzian space-time and an exponentially
damping behaviour (deflation) of $a$ for the Euclidean region.
These results are formally in general agreement with  those
obtained in \cite{EMB} and \cite{GK}. The corresponding Wheeler-DeWitt
equation is exactly solved to construct a state that may be identified with
a non-dispersive  wave packet peaking in the vicinity of the classical
Kaluza-Klein submanifold admitting signature transition. This remarkable
correspondence would help us to look at the existence of higher
dimensional geometries undergoing a continuous change of signature as
semi-classical limits in the context of quantum cosmology.

We may remark that the model presented here  does not predict the standard
big-bang but rather an eternal
cosmology.  One way of avoiding this eternal universe would be to
assume an adiabatically evolving $\Lambda$. Alternatively, one may work
within the context of a dilaton field model in which a typical
dilaton field $\phi$
with its potential $V(\phi)$ results in an effective cosmological
constant $\Lambda$. Thus, if the dilaton model admits either a duality
transformation $|\Lambda| \rightarrow |\Lambda|^{-1}$ or tunneling
from $|\Lambda|$ to zero leading to switching off $|\Lambda|$,
then it will provide for the evolution of the scale
factor $R$ from small to large scales.
Nevertheless, in its present form this eternal universe with no big-bang
and a size which is large enough (adjusting $|\Lambda|$ to a
very small constant) may have the advantage of being compatible with the
present observations and seems to be
free of the flatness problem.

It is also worth noting that if one takes a
positive cosmological constant rather than negative, one ends up
with having an inflationary rather than a compactification phase
in the Lorentzian solutions.
Finally, we remark that for (4+$D$) dimensional Kaluza-Klein
cosmology with $D>1$ in the present model, the quantization is
a difficult problem which requires further investigations.

\newpage
{\large {\bf Figure Captions}}
\vspace{10mm}
\\
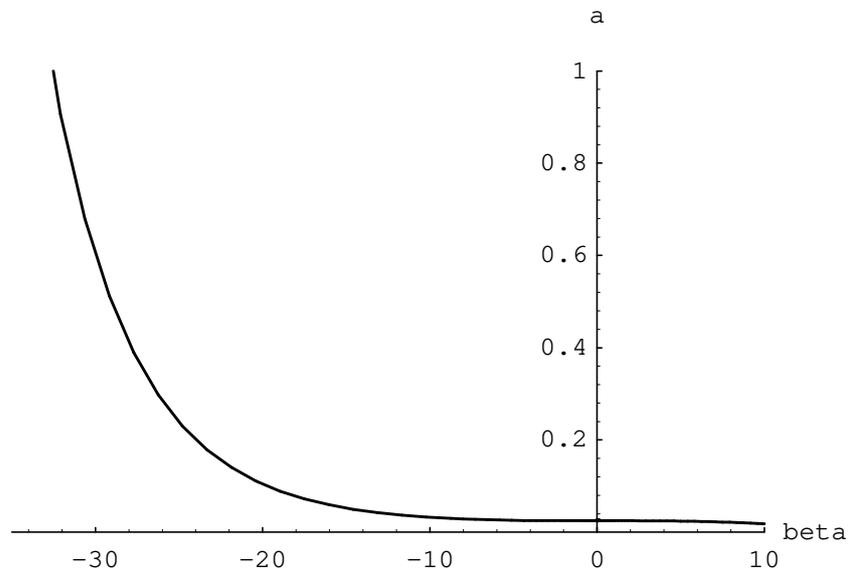
Figure 1. Variation of $a(\beta )$ with $\beta$ for a typical small
value of $\Lambda \sim 10^{-3}$.
\vspace{10mm}
\\
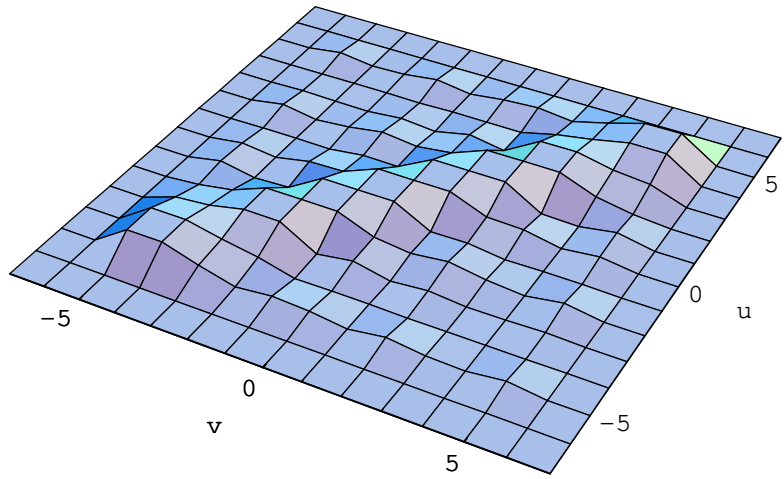
Figure 2. Surface plot of $|\Psi (u,v)|^2$.
\vspace{10mm}
\\
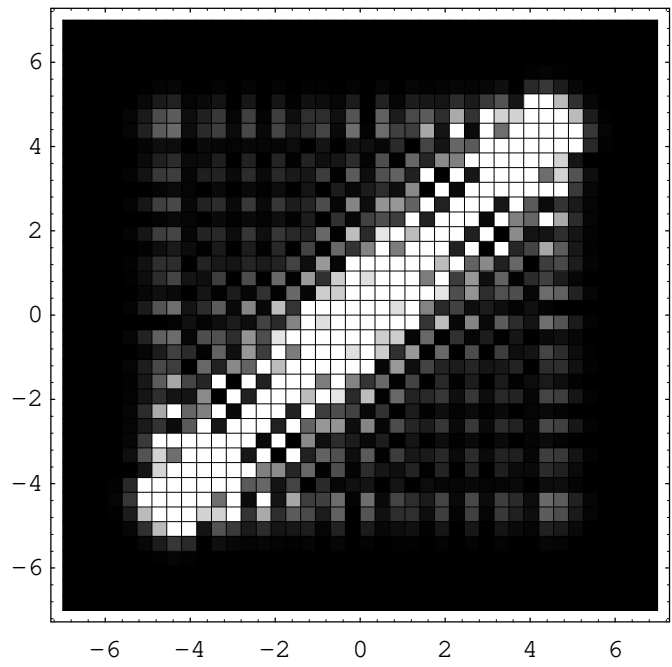
Figure 3. Density plot of $|\Psi (u,v)|^2$.

\end{document}